\documentclass[pre,twocolumn,superscriptaddress,floatfix,nofootinbib]{revtex4-1}  

\usepackage[paperwidth=210mm,paperheight=297mm,centering,hmargin=2cm,vmargin=2.5cm]{geometry}

\usepackage[pdftex]{graphicx}
\usepackage{amsmath,amsfonts,amssymb}
\usepackage{natbib}
\usepackage{MnSymbol}
\usepackage{csquotes}

\usepackage{hyperref}
\usepackage{xcolor} 
\usepackage{array}
\usepackage{pgfplots}
\usepackage{graphicx}
\usepackage[left,modulo,pagewise]{lineno}
\usepackage{dsfont}

\definecolor{bluemoi}{rgb}{0.25,0.50 ,0.75} 

\hypersetup{
  bookmarksopen=true,
  pdftitle=" ",
  pdfauthor=" ", 
  pdfsubject=" ", 
  pdftoolbar=true, 
  pdfmenubar=true, 
  pdfhighlight=/O, 
  colorlinks=true, 
  pdfpagemode=UseNone, 
  pdfpagelayout=SinglePage, 
  pdffitwindow=true, 
  linkcolor=bluemoi, 
  citecolor=bluemoi, 
  urlcolor=bluemoi 
}

\makeatletter
\renewcommand{\figurename}{\sf \textbf{Figure}}
\renewcommand{\thefigure}{\arabic{figure}}
\renewcommand{\fnum@figure}{\sf\textbf{\figurename}~\textbf{\thefigure}}
\renewcommand{\tablename}{\sf\textbf{Table}}
\renewcommand{\thetable}{\arabic{table}}
\renewcommand{\fnum@table}{\sf\textbf{\tablename}~\textbf{\thetable}}
\makeatother

\begin{document}

\title{Multiscale socio-ecological networks in the age of information}  

\author{Maxime Lenormand$^1$}
\thanks{Corresponding author: maxime.lenormand@irstea.fr\\ Authors affiliations are available at the end of the article}

\author{Sandra Luque$^1$}
\author{Johannes Langemeyer$^{2}$}
\author{Patrizia Tenerelli$^1$}
\author{Grazia Zulian$^3$}
\author{Inge Aalders$^4$}
\author{Serban Chivulescu$^5$}
\author{Pedro Clemente$^6$}
\author{Jan Dick$^7$}
\author{Jiska van Dijk$^8$}
\author{Michiel van Eupen$^{9}$}
\author{Relu C. Giuca$^{10}$}
\author{Leena Kopperoinen$^{11}$}
\author{Eszter Lellei-Kov{\'a}cs$^{12}$}
\author{Michael Leone$^{13}$}
\author{Juraj Lieskovsk{\'y}$^{14}$}
\author{Uta Schirpke$^{15}$}
\author{Alison C. Smith$^{16}$}
\author{Ulrike Tappeiner$^{17}$}
\author{Helen Woods$^{7}$}

\begin{abstract} 
Interactions between people and ecological systems, through leisure or tourism activities, form a complex socio-ecological spatial network. The analysis of the benefits people derive from their interactions with nature -- also referred to as cultural ecosystem services (CES) -- enables a better  understanding of these socio-ecological systems. In the age of information, the increasing availability of large social media databases enables a better understanding of complex socio-ecological interactions at an unprecedented spatio-temporal resolution. Within this context, we model and analyze these interactions based on information extracted from geotagged photographs embedded into a multiscale socio-ecological network. We apply this approach to 16 case study sites in Europe using a social media database (Flickr) containing more than 150,000 validated and classified photographs. After evaluating the representativeness of the network, we investigate the impact of visitors’ origin on the distribution of socio-ecological interactions at different scales. First at a global scale, we develop a spatial measure of attractiveness and use this to identify four groups of sites. Then, at a local scale, we explore how the distance traveled by the users to reach a site affects the way they interact with this site in space and time. The approach developed here, integrating social media data into a network-based framework, offers a new way of visualizing and modeling interactions between humans and landscapes. Results provide valuable insights for understanding relationships between social demands for CES and the places of their realization, thus allowing for the development of more efficient conservation and planning strategies.
\end{abstract}

\maketitle

As visitors’ priorities and consumption patterns evolve, people are travelling more frequently, further away from home, and in greater numbers \cite{UNWTO}. People interact with the destination sites, affecting landscapes, societies and quality of life. Hence, these recent changing mobility patterns open up new challenges in understanding threats and constraints to the environment. Leisure or tourism activities affect cities and their surroundings, as well as remote natural areas, through the impact of travel movements and the presence of people \cite{Levin2015,Schirpke2018}. Socio-ecological interactions generate, in turn, cultural ecosystem services (CES) and relational values, linking people and ecosystems via tangible and intangible relationships \cite{Cao2015}. Visitors move according to personal preferences, often influenced by the attractiveness of an area. To gain an understanding of visitor patterns and how humans interact with their environment, it is essential to undertake a holistic approach to socio-ecological systems, by focusing on the different components of the system and the way they interact with each other. Models of spatial relations between CES realization areas and beneficiaries based on empirical data are needed to disentangle interdependencies between social and ecological systems at a high spatio-temporal resolution.

A promising approach is to consider socio-ecological systems as networks \cite{Knights2013}. Indeed, nature-based interactions can be represented as a spatial network \cite{Barthelemy2011} that offers a way of visualizing and analyzing multiscale spatio-temporal CES demands linked to a particular site. However, the lack of data represents an important limitation for the modeling of CES emerging from socio-ecological interactions particularly at a global scale. Traditional data sources such as census or surveys usually fail at mapping human population dynamics during situations in which detailed spatio-temporal information is required \cite{Deville2014}, as in the analysis of individual human spatio-temporal trajectories. ICT devices such as mobile phones are now widely accessible and generate a large quantity of high resolution spatio-temporal information on individual human mobility patterns \cite{Brockmann2006,Gonzalez2008,Song2010,Lenormand2015,Barbosa2018}. The reliability and the accuracy of these new data sources have been intensively evaluated in recent years, notably by comparing mobility information extracted from ICT data and more traditional data sources \cite{Deville2014,Tizzoni2014,Lenormand2014,Lenormand2016}. Among these new data sources, of particular interest is geotagged information produced via social media that has been increasingly used in many scientific fields to study human mobility patterns \cite{Barbosa2018}. Among the most popular, Twitter data has been widely used in understanding social networks \cite{Java2007,Krishnamurthy2008,Grabowicz2012} and how people interact with the built environment \cite{Frias2012,Hawelka2014,Lenormand2014,Lenormand2015}. Data retrieved from the Flickr photo-sharing platform have been notably used for the identification of users' home locations \cite{Bojic2015} and the modelling of individual human mobility patterns \cite{Barchiesi2015}. Nevertheless, these studies usually focus on the way people interact with each other and with their environment in urban systems. More recently, the digital traces that we leave while visiting touristic and natural spaces have also contributed to the assessment of cultural ecosystem services \cite{Langemeyer2018,Schirpke2018,Casalegno2013,Tenerelli2016, Figueroa2017}, the measurement of landscape values \cite{Martinez2016,Zanten2016}, the attractiveness of tourist sites \cite{Wood2013,Bassolas2016,Tenerelli2017} and the monitoring of visitors in protected areas \cite{Levin2015,Levin2017,Tenkanen2017}. These studies represent a crucial step towards a better understanding of interactions between people and ecological systems, through leisure or tourism activities, but they usually focus only on the presence of individuals on a site, and do not explicitly take into account the spatial relation between humans and nature that underlies beneficial socio-ecological interactions in situ, nor information about the individuals that visit a site.

The aim of this paper is to explore the potential of Flickr data for the study of socio-ecological interactions. The guiding idea is that interactions between individuals and ecological systems can be visualized and modeled using geotagged photographs from the Flickr photo-sharing platform embedded in a multiscale socio-ecological network. Based on more than $150,000$ photos taken in 16 study sites across Europe, this study examines the potential of the digital traces that we leave while visiting natural sites to efficiently represent socio-ecological interactions at different scales. 

\section*{Results}

\subsection*{Extracting a multiscale socio-ecological network from social media data}

\begin{figure}[!ht]
	\centering 
	\includegraphics[width=\linewidth]{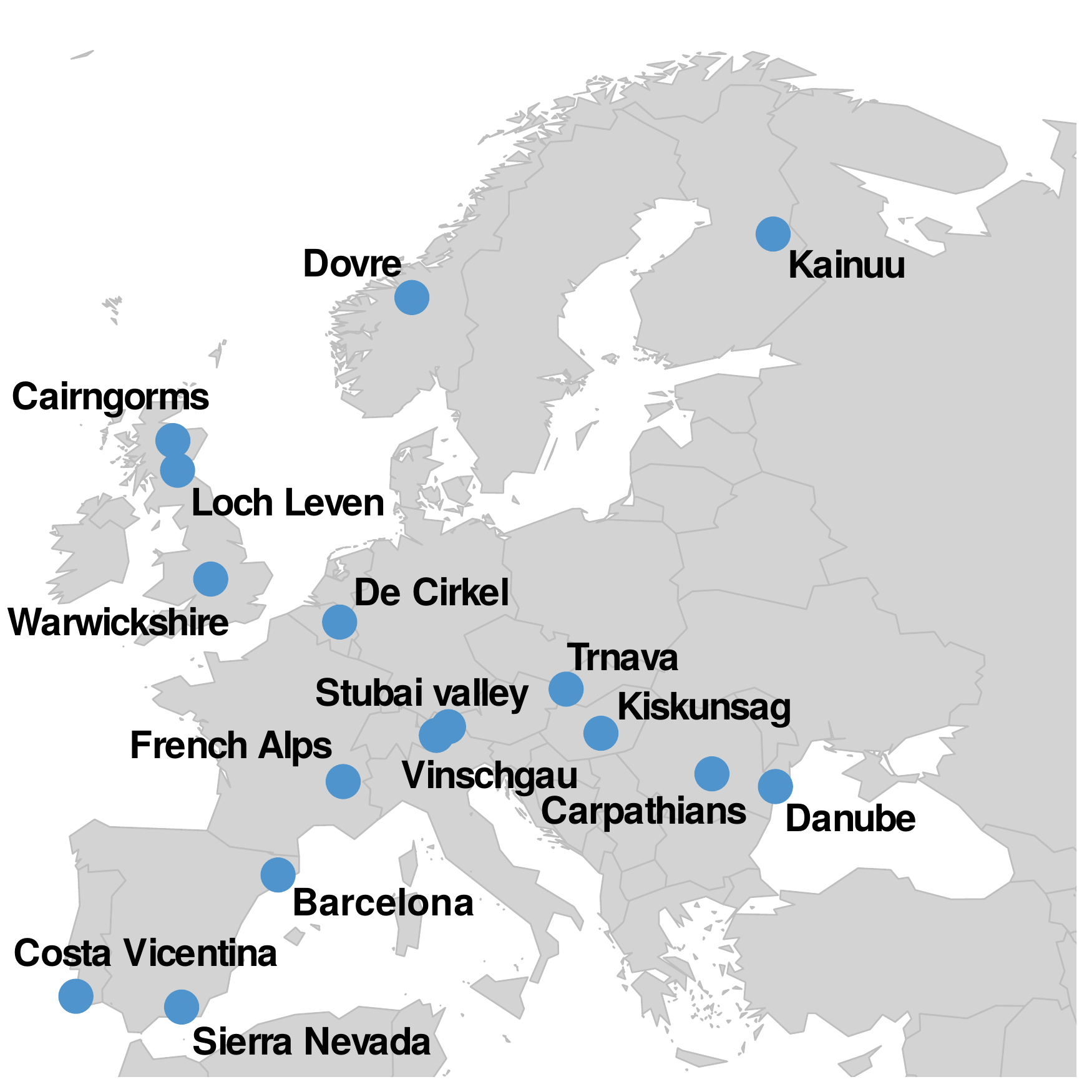}
	\caption{\textbf{Positions of the 16 case study sites.} \label{Fig1}}
\end{figure}

\begin{figure*}[!ht]
	\centering 
	\includegraphics[width=\linewidth]{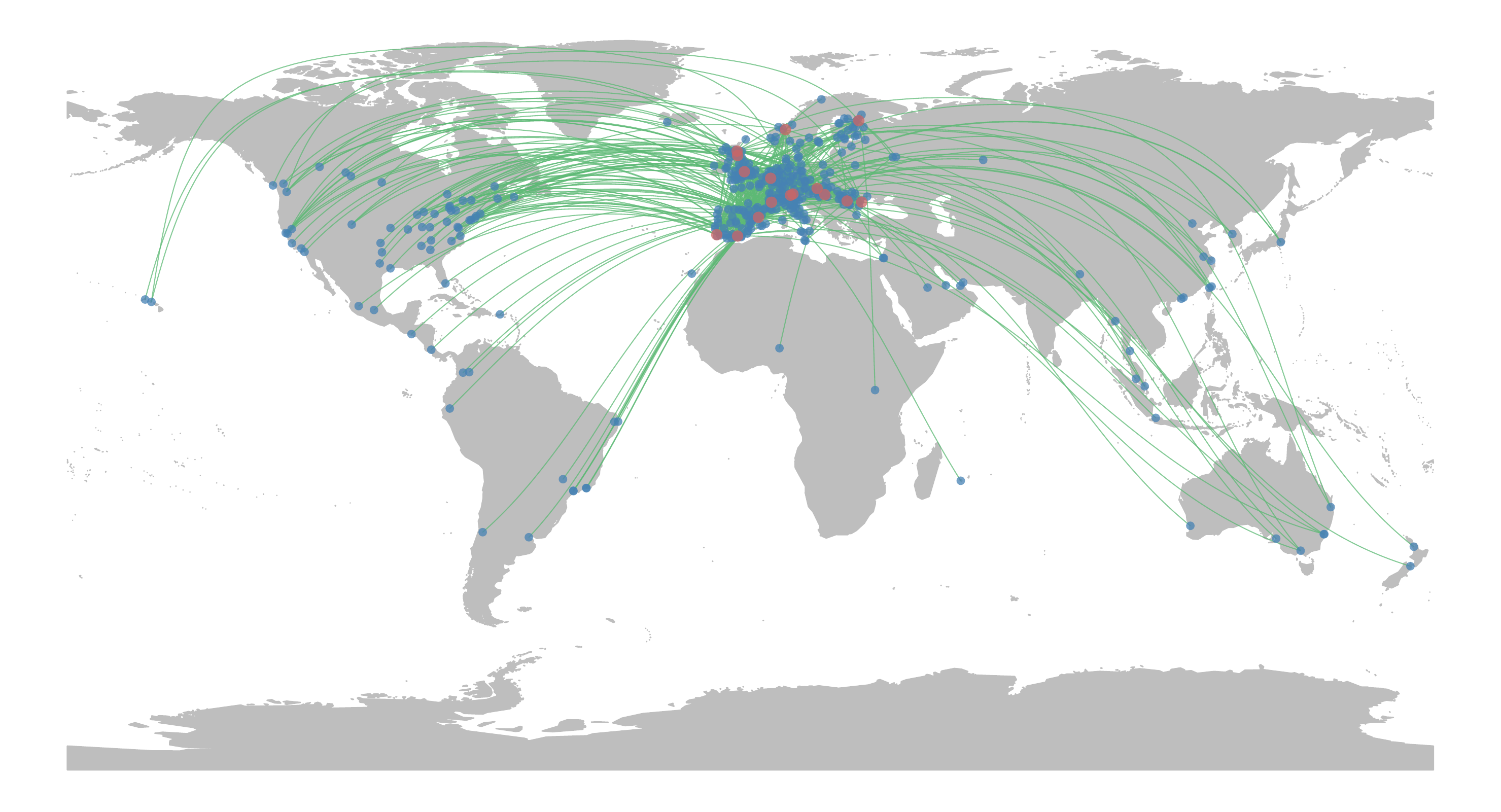}
	\caption{\textbf{Representation of the socio-ecological network at a global scale.} Every users' place of residence (blue dots) are linked to the case study sites (red dots) by one or more interactions (green curves). \label{Fig2}}
\end{figure*}

Socio-ecological interactions have been extracted from a database containing more than $150,000$ photographs taken between 2000 and 2017 in 16 sites in Europe (Figures \ref{Fig1} and \ref{Fig2}) and posted on the Flickr social media platform. Each photo is geo-localized (latitude/longitude coordinates), time-stamped and associated with a unique Flickr user ID. In order to ensure that only photographs representing an interaction between an individual and a natural site are considered, each photo has been manually validated and classified according to the landscape and the activity identified on the picture. These validated photographs have been taken and posted on Flickr by $2,193$ reliable users whose place of residence have been identified based on their Flickr timeline using $100 \times 100 \mbox{ km}^2$ world grid cells. See the Materials and Methods section for more details. We define a socio-ecological interaction as the presence of a Flickr user in one of the 16 sites during a given time window. The individuals are characterized by their place of residence. The ecological systems are represented by a geographical location at different scales. Two scales are considered, a global scale (16 European sites) and a local one where every site has been divided in zones using $500 \times 500 \mbox{ m}^2$ grid cells. In practice, an interaction is represented by one or several photos taken by a user in a grid cell during a given hour. Note that if several photos are taken during an interaction, the different types of interactions (landscapes and activities) identified on the photos are aggregated. The resulting network is composed of $7,354$ socio-ecological interactions linking $365$ distinct places of residence all over the world to $3,418$ grid cells located in 16 study sites. A spatial representation of the network at a global scale is displayed in Figure \ref{Fig2}. 

\subsection*{Evaluation of the network's representativeness}

\begin{figure}[!ht]
	\centering 
	\includegraphics[width=\linewidth]{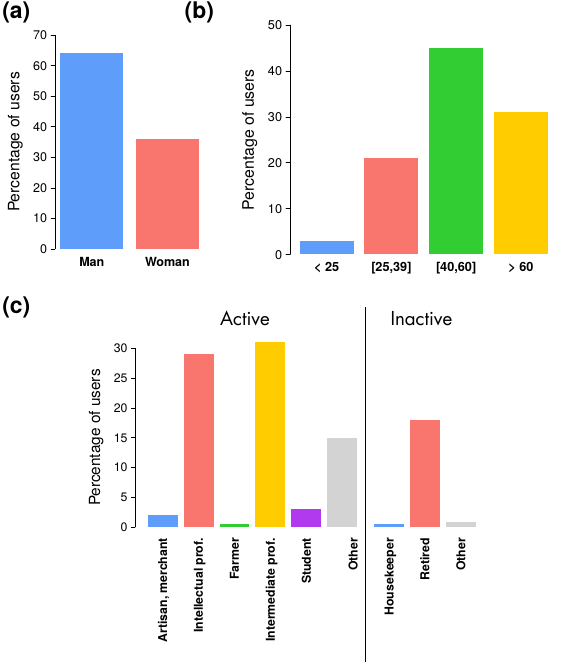}
	\caption{\textbf{Results of the survey.} Percentage of respondents according to the gender (a), age (b) and socio-professional category (c). \label{Fig3}}
\end{figure}

New data sources such as Flickr data have the great advantage of being global, in contrast with surveys and census data involving usually only one country or at most only a few countries. In return, they come with several biases associated with the lack of information regarding the users’ sociodemographic characteristics. In order to collect more information about our sample we automatically sent a questionnaire to the $2,193$ reliable users of our cleaned database through the creation of a Flickr group. We obtained a response rate of 11\%. Figure \ref{Fig3} shows some descriptive statistics about the respondents according to their socio-demographic characteristics. We note that men represent about two thirds of the respondents. There are also very few young people, the respondents were predominently professionals. By asking the respondents to provide us with their zipcode and country of residence, this survey supported the identification of the user's place of residence based on the Flickr timeline (more details available in the Materials and Methods section and in Appendix). The overall agreement is good: in 90 percent of cases, the location entered in the questionnaire is located within the $100 \times 100 \mbox{ km}^2$ world grid cell detected with our algorithm. 

\subsection*{Sites' attractiveness}

\begin{figure*}[!ht]
	\centering 
	\includegraphics[width=16cm]{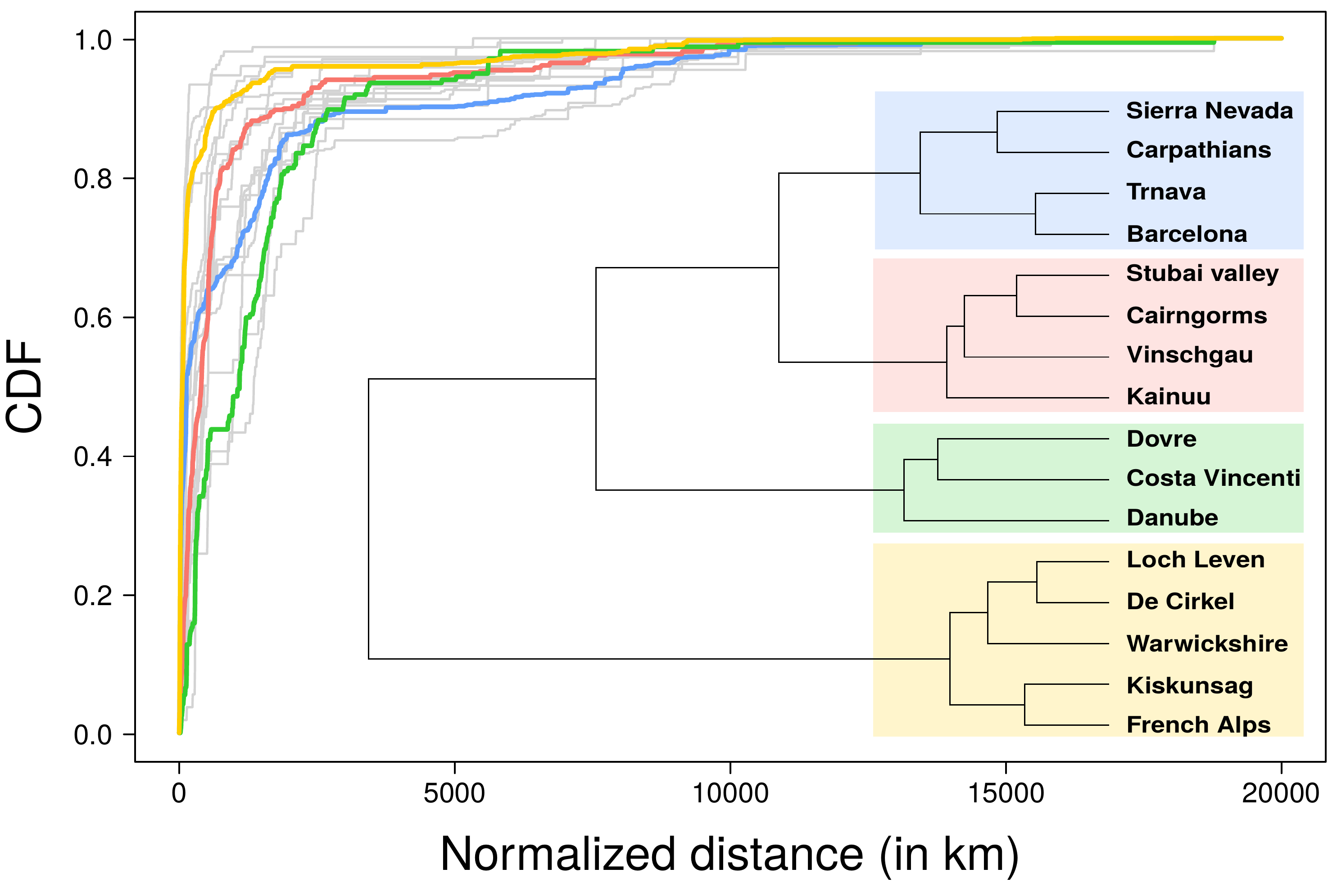}
	\caption{\textbf{Measure of the sites' attractiveness.} Cumulative distribution function (CDF) of the normalized distance between users' places of residence and case study sites. Each grey curve represents a case study. Four common profiles were found using ascending hierarchical clustering (AHC). Each colored curve represents one of this profile (average CDF in each cluster). The dendrogram resulting from the hierarchical clustering is shown in inset. \label{Fig4}}
\end{figure*}

Being able to measure quantitatively the interactions between a particular site and the rest of the world allows for the development of attractiveness indicators that have been already successfully applied to cities \cite{Lenormand2015} or touristic sites \cite{Bassolas2016,Tenerelli2017} in the past. Among these metrics, of particular interest is the average distance traveled by the visitors to reach the site. Figure \ref{Fig4} displays the Cumulative Distribution Function (CDF) of the normalized distance between Flickr users' place of residence and case study sites in our network. Note that to take into account the sites' accessibility, the distance was normalized beforehand (see the Materials and Methods section for more details). The global attractiveness of a site can be inferred from the area above the curve, while the shape of the curve informs us on the type of attractiveness. We observe that some case study sites are more attractive than others, highlighting different levels of attractiveness from local to global influence. The hierarchical cluster analysis using the Ward distance identified similarities between CDFs. Four well-separated clusters are identified; the corresponding average CDFs are represented by the colored curves in Figure \ref{Fig4}. The yellow cluster is composed of case study sites having a local influence, while the other case studies tend to attract people coming from further away. Key examples in mountain regions are the Vercors in the French Alps, that is mainly visited by locals and from nearby cities (yellow cluster), while the Sierra Nevada in Spain and the Carpathians have an international reputation and share the blue cluster.  Sites composing the blue and green clusters have a high level of attractiveness, at a more global level for the blue one than for the green one. 

\subsection*{Effect of the distance traveled on the socio-ecological interactions}

To evaluate how the distance to a site influences the way people explore and interact with this site, we apply five metrics. These metrics summarize the distribution of interactions from a spatial and a temporal dimension, but also from the point of view of landscape diversity. We analyze basic characteristics of the spatial distribution of interactions taking into account the spatial coverage (number of cells with at least one interaction), the spatial dispersion of interactions in the cells measured with an entropy index, and a spatial dilatation index measured as the average distance between interactions. We also measure the temporal dispersion of visits throughout the year at a monthly granularity. Finally, we assess landscape diversity using six landscape categories. More details on the metrics are given in the Materials and Methods section.

\begin{figure}[!ht]
	\centering 
	\includegraphics[width=\linewidth]{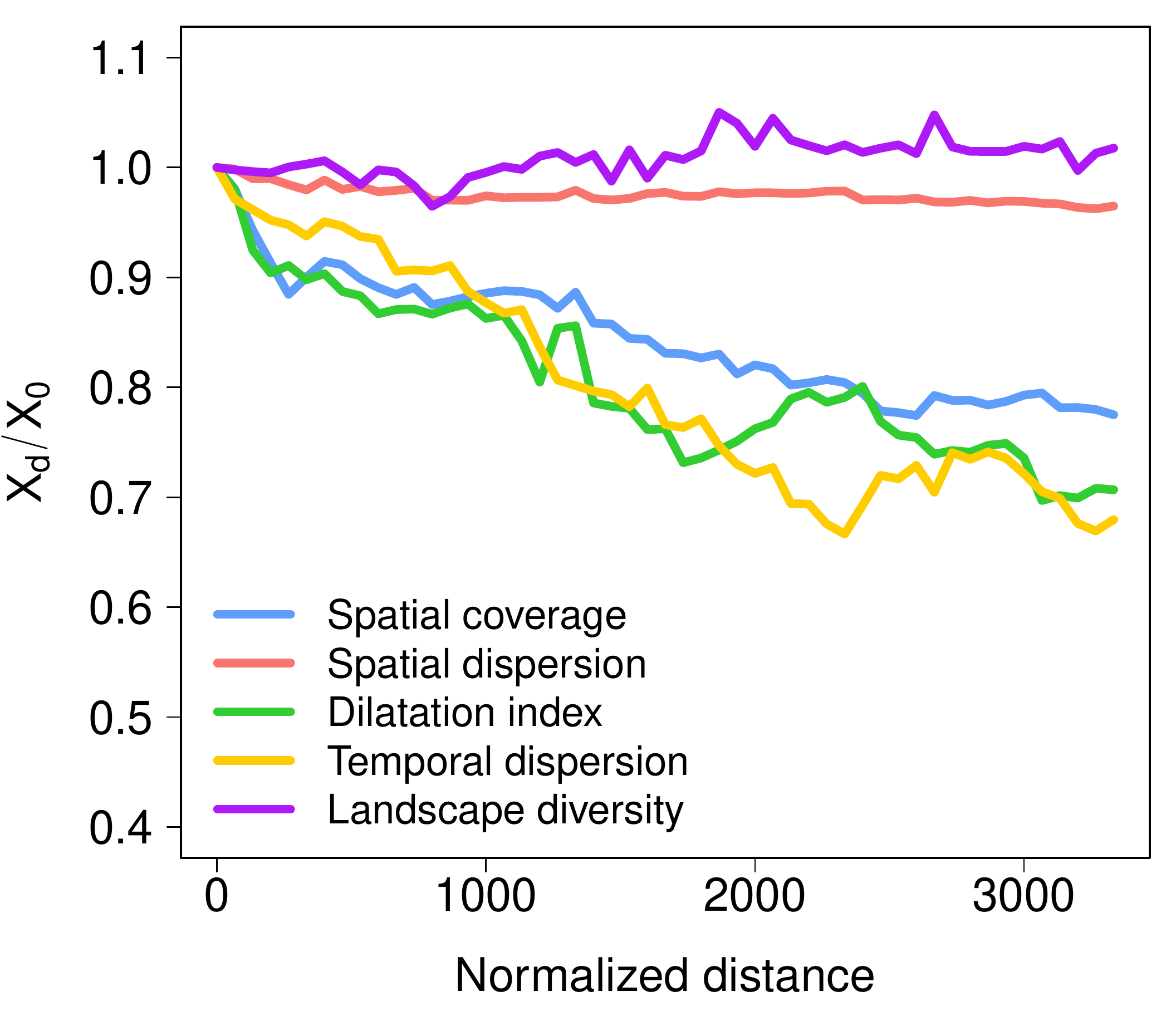}
	\caption{\textbf{Effect of the distance traveled on the socio-ecological interactions.} Evolution of the spatial coverage (blue), the spatial dispersion (red), the spatial dilatation index (green), the temporal dispersion (yellow) and the landscape diversity (purple) as a function of the normalized distance. For each metric, the median over the 16 case studies is displayed. All metrics are normalized by the value obtained with a random null model. Similar plots for each case study are available Figure S7 in Appendix. The effect of the spatial resolution on the spatial metrics is presented in Figure S8. \label{Fig5}}
\end{figure} 

In order to assess the effect of the distance traveled we compare the results obtained considering only interactions made by individuals living further than a certain normalized distance to the ones obtained under the null hypothesis that does not take into account the distance, considering therefore all the interactions. However, most of the metrics used are affected by the sample size (i.e. number of interactions). To side-step this difficulty, we introduce a random null model accounting for the distribution with different sample sizes. The five  metrics computed as a function of the normalized distance are plotted in Figure \ref{Fig5}. Each point on the curve represents the value of a metric $X_d$, taking into account the normalized distance traveled $d$, divided by $X_0$, the value obtained with a random null model assuming that the distance has no influence on the metric (more details in Materials and Methods). We observe that the area covered by the interactions and the dilatation index decrease with the distance traveled, while the interactions tend to be spatially distributed in a similar way whatever the distance traveled (as measured with the spatial dispersion index). Hence, as the distance traveled increases, the visitors tend to explore the area less, though the pattern of dispersal within the space explored is similar. In contrast, regarding the temporal aspect of the distribution, we observe that the interactions tend to be more concentrated in a certain period of the year as the distance traveled increases. Finally, it is interesting to note that the complexity of interactions in terms of landscape diversity increases relatively little with the traveled distance. However, it is important to keep in mind that these observations represent a median behavior across the 16 case study sites and are not always representative of all case studies, particularly regarding the landscape diversity metric (see Figure S7 in Appendix).

\begin{figure*}[!ht]
	\centering 
	\includegraphics[width=\linewidth]{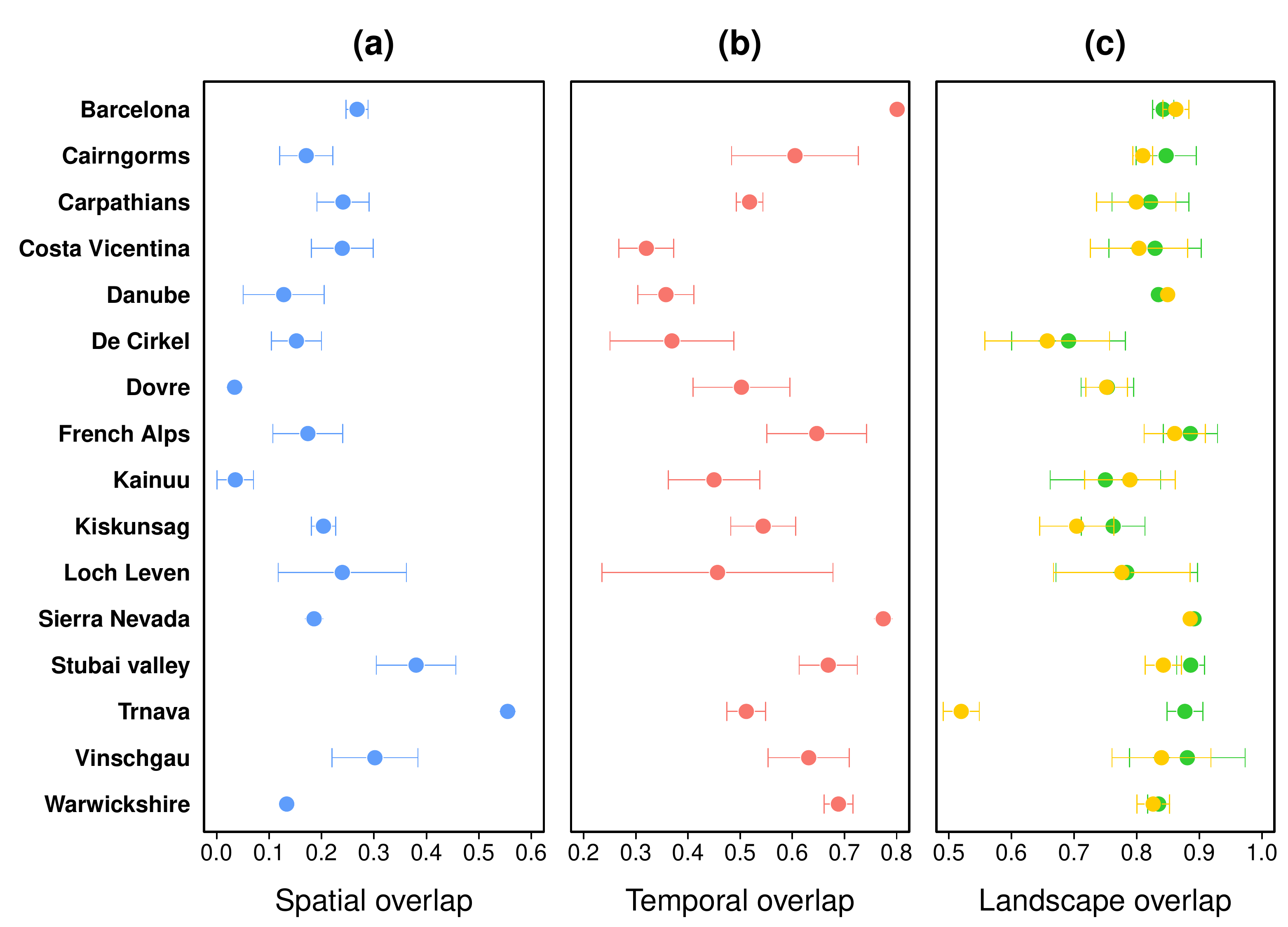}
	\caption{\textbf{Overlap between locals and visitors' interactions.} Spatial (a), temporal (b) and landscapes (c) overlap between locals and visitors' interactions. In panel (c), the green points represent the landscape overlap between locals and visitors considering all the cells, while the yellow points represent the landscape overlap between locals and visitors in cells frequented exclusively by locals from one side and visitors from the other side (without spatial overlap). Locals and visitors are identified according to the normalized distance. In order to assess the impact of the threshold on the results we averaged the metrics obtained with threshold values ranging between $100$ and $1,000$ km. The error bars represent one standard deviation. The effect of the spatial resolution on the spatial overlap is presented in Figure S9 in Appendix. \label{Fig6}}
\end{figure*}

\subsection*{Locals and visitors' interactions overlap}

We focused so far on the influence of the distance traveled by the users on the distribution of their socio-ecological interactions. However, an important question remains: does the pattern and intensity of the interactions depend on the origin of the user? To answer this question, in this section we analyze the overlap between locals’ and visitors' interactions. The interactions are first separated into two groups according to the users' place of residence. A user is considered as local if the normalized distance between her/his place of residence and the site is lower than a predetermined threshold; otherwise they are considered as a visitor. The overlap is defined as the fraction of interactions in common between locals’ and visitors’ distributions depending on the dimension being considered (spatial, temporal or based on landscape diversity). Here again, to make the locals’ and visitors’ distributions comparable, we use a random model taking into account the difference in sample size (see the Materials and Methods section for more details). To assess the impact of the threshold used to separate the two groups of users, the results have been aggregated over different threshold values ranging between $100$ and $1,000$ km. Figure \ref{Fig6} shows the average and standard deviation obtained for each dimension. The spatial overlap between visitors’ and locals’ interactions is relatively low, with values fluctuating around 25\% of overlap between the two spatial distributions. Figure \ref{Fig6}b shows the temporal overlap between locals’ and visitors' interactions: although the results are more heterogeneous, the overlap is globally higher but still quite low with an average overlap of 50\%. These results tend to demonstrate that locals and visitors interact with natural spaces differently in space and time. This is less true for the type of landscapes observed during the interactions, with an 80\% overlap between locals and visitors (green points in Figure \ref{Fig6}c). It must be noted that focusing only on the type of landscapes observed by locals and visitors in cells frequented only by locals and visitors (i.e. without spatial overlap) does not significantly change the results except for the site of Trnava (yellow points in Figure \ref{Fig6}c).

\section*{Discussion}

Central to the measure of human perception and interest in natural environments is the concept of CES, but it is challenging to relate the supply of these non-material services to specific spatial units. Moreover, the attractiveness of a site and the way we explore it may be influenced by our origins \cite{Chan2016, Tenerelli2017}. Indeed, beyond the analysis of people’s activities in natural sites, an important question remains: how does our origin impact the nature of our relationships with natural ecosystems? Taking advantage of a social media database, we proposed in this work a methodological approach to extract and analyze multiscale socio-ecological networks from volunteered, publicly available data generated from social media. We extracted and analyzed from a Flickr database $7,354$ socio-ecological interactions made in 16 case study sites in Europe by individuals living all around the world. Two scales have been considered. First a global scale, focusing on the sites’ attractiveness based on the distance traveled by the users, and then a local scale, by analyzing how the way Flickr users explore a site varies with the distance traveled to reach this site.

Our results demonstrate that while different levels of attractiveness exist among sites (local, regional and global), the existence of differences in the patterns of socio-ecological interactions according to visitors' origins is remarkably consistent across sites. Indeed, the distance traveled has a significant effect on the way Flickr users interact with natural ecosystems in both the spatial and temporal dimensions. Although further research in this direction is needed, it would appear that the desire for landscape diversity in socio-ecological interactions does not vary significantly with the distance traveled to reach a site. Of particular interest is the concept of overlaps between locals and visitors that could be used within the framework of planning strategies oriented towards conservation and sustainable tourism, for example to improve management of visitor activities in protected areas in order to reduce human impacts \cite{Schulze2018}.

\subsection*{Limitations of the study}

In this work, we explore the possibility of making use of social media data to provide information about the way people interact with ecological systems. In particular, we developed a methodology to connect Flickr users' place of residence to places where the interaction took place in the 16 study sites. This allowed us to study how the distribution of interactions varies among sites and according to the distance traveled. Although it would not have been possible to conduct this research at global scale using conventional data sources such as surveys, it is necessary to recognize the potential limits and biases of our approach.

First, we cannot ensure the reliability of the data both in terms of space and time resolution. In order to limit the potential biases we considered the photos with the most precise spatio-temporal Flickr accuracy level (according to the Flickr API which gives access to the accuracy with which Flickr knows the date and location to be registered). It is also worth noting that most of our analysis is based on data aggregated both in space ($100 \times 100 \mbox{ km}^2$ world grid cells and $500 \times 500 \mbox{ m}^2$ case study site grid cells) and time (month granularity). The spatial aggregation allowed to overcome the spatial accuracy error linked to the used GPS-enabled devices, or the map scale used to specify the photo location. By using a manual photograph validation and classification process, we were able to avoid potential errors in classification that arise when using automatic image processing tools. However, even though the interpretation of the photographs was performed by between 1 and 6 local experts following a rigorous protocol, interpretation of the images may still be subjective to some extent. 

Another important limitation lies in the lack of information regarding the characteristics of individuals using Flickr. The process of identifying the user's place of residence allows us to discard non-reliable Flickr users (those with a collective account or who are not regular Flickr users). A first coarse filter was applied to exclude collective accounts from the data. Then, we applied several filters to ensure that a user shows enough regularity and that the assigned place of residence is the region of the world where he/she is really living (see the Appendix for more details). Nevertheless it was also important for us to be able to evaluate the performance of our place of residence detection algorithm. This is why we decided to integrate an online survey in our analysis. Although the response rate was quite low (11\%), this survey permitted us to get a better understanding of the sociodemographic characteristics of Flickr users, which is usually an important limitation of this kind of study. We believe that the integration of online survey approaches combined with crowdsourced data might overcome some of the limitations of using geotagged public photos to analyze the way people interact with nature. 

All these filters tended to reduce the size of our initial sample. This severely limited the possibility of performing multi-dimensional analysis (considering space, time and landscape diversity at the same time). Nevertheless, we rigorously studied these dimensions separately and to limit sample size effects we have introduced null models taking into account the sample size and its variability.

Finally, since the distance between a user's origin and the site visited can be biased by the geography, we took this heterogeneity into account by measuring the case study sites' accessibility. In this process, all the distances have been computed with the Haversine formula based on longitude and latitude coordinates. However, distances as the crow flies are rarely a direct proxy for travel time particularly at local scale. In future studies, flight distances, transport APIs and road network data could be used instead to calculate more realistic travel distances between different points on the globe.

\subsection*{Concluding remarks}

Within the framework of this research we also developed a visualization application to provide stakeholders with a tool based on the analysis that could be used for planning (more details in Appendix). This web application is also oriented towards Flickr users who participated actively (providing input via the survey) or passively to the experience, and could become a platform in the future to share experiences from the photos and the visual content. Such a platform could limit the biases mentioned above, allowing the users to classify their own photos supported with image processing tools and to fill in an anonymous online survey improving knowledge about their origin and motivations. 

Hence, following the approach proposed in this paper, further studies could consider the sociodemographic characteristics as well as psycho-cultural aspects which could reveal significant correlation with the knowledge or appreciation of specific ecosystems. Indeed this approach opens the door to future analysis and applications; further investigation is certainly needed to understand complex human-ecosystem interactions.

\section*{Material and Methods}

\subsection*{Photograph classification process}

In order to ensure that only photographs representing an interaction between an individual and an ecosystem are considered, the subject of each photo was manually validated and classified according to the landscape identified in the picture and the different types of cultural services that people benefits from ecosystems. In this study, we focused on six landscape categories: agricultural and open landscape, sparse forest landscape, forested landscape, mountain landscape, manmade infrastructure, water landscapes and wetlands. At the end of the process, $16,716$ photos taken by $2,967$ users between January 2000 and 2017 were classified. Note that 98\% of the photos were taken after 2007. More details about the photograph classification process are available in Appendix.

\subsection*{Identification of the user's place of residence}

To identify the place of residence of the $2,967$ Flickr users, we retrieved through the Flickr API information related to all the geo-located photos taken by these users worldwide. Then, we divided the world using a grid composed of $100 \times 100$ square kilometer cells in a cylindrical equal-area projection. We define a user’s place of residence as the cell from which she or he has taken most of her/his photos \cite{Lenormand2016}. After discarding users where the place of residence could not be identified, we obtained $12,850$ classified photos taken by $2,193$ users between January 2000 and January 2017 in the 16 case studies. More details about the method are available in Appendix.

\subsection*{Accessibility and attractiveness}

For each user, we computed the distance between their place of residence and the centroid of the study site he or she has visited. Since the distance between the user’s origin and the visited site can be biased by the geography, we computed a normalized distance taking into account the origin of the user and the accessibility of the site. All the distances have been computed with the Haversine formula based on longitude and latitude coordinates.  More details about the method used to compute the sites’ accessibility are available in Appendix. 

\subsection*{Description of the metrics}

\subsubsection*{Spatial dimension}

In order to investigate the impact of the distance traveled on socio-ecological interactions, we defined a set of metrics to characterize them. We focus here on the spatial and temporal dimensions, but also on the diversity of landscapes identified in the photos. Three indicators are used to characterize the spatial distribution of interactions. The spatial coverage is defined as the area covered by the socio-ecological interactions, estimated as the number of $500 \times 500 \mbox{ m}^2$ cells in which at least one interaction occurred. This metric does not take into account the density of interactions in each cell nor the morphology of the spatial distribution. To compensate these limitations we also introduced a spatial dilatation index defined as the average distance between all the interactions and a metric of spatial dispersion to evaluate whether the interactions are concentrated in a few cells or evenly distributed within the surface covered by the interactions. We measure the spatial dispersion with a spatial entropy index. If we define the probability $p_i$ that an interaction occurs in a cell $i$, then the entropy $E$ is given by:
\begin{equation}
E=-\frac{\sum_{i} p_i\log(p_i)}{A}  
\label{E}
\end{equation}
where the normalizing factor $A$ is equal to the number of cells with at least one interaction. A value close to 0 means that the majority of the interactions are clustered in a few cells, and al value close to 1 indicates that the interactions are uniformly distributed among the cells.

\subsubsection*{Temporal dimension}

To get a better understanding of the way socio-ecological interactions are distributed within the year, and whether or not the distance traveled affects this distribution, we also rely on the entropy index to compute the temporal dispersion of interactions. In this case, we define the probability $p_i$ that an interaction occurs during a given month and the normalizing factor $A$ is equal to $log(12)$. 

\subsubsection*{Landscape diversity}

Another important dimension to consider is the diversity of landscapes present in the photographs. Here again, the landscape diversity is based on an entropy metric considering the probability $p_i$ to interact with particular landscape categories (agricultural and open landscape, sparse forest landscape, forested landscape, mountain landscape, manmade infrastructure or water landscapes and wetlands). Each interaction is characterized by a vector representing the probability to interact with the six landscape categories. The entropy is computed as an average over all the considered interactions. In this case, the normalizing factor $A$ is equal to $log(6)$.

\subsubsection*{Overlap}

The way socio-ecological interactions are distributed in space, time or type of landscapes may depend on the origin of the user. To answer this question we analyze the overlap between locals’ and visitors' interactions. We define the overlap between two distributions of probability $p$ and $q$ on the same finite support as follows,
\begin{equation}
O=\sum_{i} \min(p_i,q_i)   
\label{O}
\end{equation} 
The distribution of probabilities $p$ and $q$ can be based on the fraction of locals’ and visitors’ interactions per cell (spatial dimension), month (temporal dimension) or type of landscapes (landscape diversity).

\subsection*{Null models}

In order to assess the effect of the distance traveled on the metrics described above, we need to compare their values to the ones returned by a random null model that does not take into account the distance. Each indicator $X$ described above can be calculated from a distribution of interactions considering only users living at a normalized distance higher than $d$ from the sites. To be meaningful, this new indicator value $X_d$ that takes into account the distance needs to be normalized by $X_0$, the value obtained with a random null model that does not take into account the distance and based on the same number of interactions. More specifically, $X_0$ is computed with the same number of interactions as $X_d$, drawn at random among all the interactions without taking into account the distance. The value of $X_0$ is averaged over $100$ replications. 

Regarding the comparison between locals’ and visitors’ interactions, the two distributions are made comparable by taking the distribution with the lowest number of interactions as a reference, and drawing at random the same number of interactions in the second distribution to obtain a distribution of the same size. The overlap between these two distributions is then computed and averaged over $100$ replications. 

\section*{Authors contribution}

ML designed the study, processed and analyzed the data, and wrote the paper. SL coordinated the study. SL, PT, GZ, IA, SC, PC, JD, JD, ME, LK, JL, ELK, ML, JL, US, AS, UT and HW coordinated the manual photographs validation and classification process in each case study. All authors read, commented and validated the final version of the manuscript.

\section*{Acknowledgments}

This publication has been funded by the ALTER-Net network through the AHIA program. This study was partially supported by the FP7 OpenNESS project (grant agreement 308428). ML thanks the French National Research Agency for its financial support (project NetCost, ANR-17-CE03-0003 grant).

\bibliography{FlickrNetwork}

\vspace*{2cm}
\onecolumngrid
\section*{Authors affiliations}

\noindent Maxime Lenormand$^1$, Sandra Luque$^1$, Johannes Langemeyer$^{2}$, Patrizia Tenerelli$^1$, Grazia Zulian$^3$, Inge Aalders$^4$, Serban Chivulescu$^5$, Pedro Clemente$^6$, Jan Dick$^7$, Jiska van Dijk$^8$, Michiel van Eupen$^9$, Relu C. Giuca$^{10}$, Leena Kopperoinen$^{11}$, Eszter Lellei-Kov{\'a}cs$^{12}$, Michael Leone$^{13}$, Juraj Lieskovsk{\'y}$^{14}$, Uta Schirpke$^{15}$, Alison C. Smith$^{16}$, Ulrike Tappeiner$^{17}$ and Helen Woods$^{6}$\\

\vspace*{0.5cm}

\noindent $^1$ Irstea, UMR TETIS, 500 rue JF Breton, FR-34093 Montpellier, France\\

\noindent $^2$ Institute of Environmental Science and Technology, Universitat Autònoma de Barcelona,  C/ de les Columnes s/n, Campus UAB, 08193 Bellaterra, Spain\\

\noindent $^3$ European  Commission,  Joint Research Centre (JRC), Directorate D - Sustainable Resources, Unit D3 - Land Resources, Ispra, Italy\\

\noindent $^4$ The James Hutton Institute, Craigiebuckler, Aberdeen, AB15 8QH, UK\\

\noindent $^5$ National Institute for Research and Development and Forestry, Blvd. Eroilor 128, 077191, Voluntari, Ilfov, Romania\\

\noindent $^6$ Center for Environmental and Sustainability Research (CENSE), NOVA School of Science and Technology NOVA University Lisbon, Campus da Caparica, 2829-516, Caparica, Portugal\\

\noindent $^7$ Centre for Ecology \& Hydrology, Bush Estate, Penicuik, EH26 0QB, UK\\

\noindent $^8$ Norwegian Institute for Nature Research (NINA), H{\o}gskoleringen 9, 7034 Trondheim, Norway\\

\noindent $^{9}$ Wageningen University and Research, Environmental Research, P.O. Box 47, 6700 AA Wageningen, The Netherlands\\

\noindent $^{10}$ Research Center in Systems Ecology and Sustainability, University of Bucharest, Splaiul Independentei 91-95, 050095, Bucharest, Romania\\

\noindent $^{11}$ Finnish Environment Institute, P.O.Box 140, FI-00251 Helsinki, Finland\\

\noindent $^{12}$ Institute of Ecology and Botany, MTA Centre for Ecological Research, Alkotm{\'a}ny u. 2-4., 2163-V{\'a}cr{\'a}t{\'o}t, Hungary\\

\noindent $^{13}$ Research Institute for Nature and Forest (INBO), Havenlaan 88 bus 73, 1000 Brussels, Belgium\\

\noindent $^{14}$ Institute of Landscape Ecology, Slovak Academy of Sciences, Akademick{\'a} 2, 949 01 Nitra, Slovakia\\

\noindent $^{15}$ Institute for Alpine Environment, Eurac Research, Viale Druso 1, 39100 Bolzano, Italy\\

\noindent $^{16}$ Environmental Change Institute, University of Oxford, Dyson Perrins Building, South Parks Road, Oxford OX1 3QY, UK\\

\noindent $^{17}$ Department of Ecology, University of Innsbruck, Sternwartestr. 15, 6020 Innsbruck, Austria\\

\newpage
\twocolumngrid

\makeatletter
\renewcommand{\fnum@figure}{\sf\textbf{\figurename~\textbf{S}\textbf{\thefigure}}}
\renewcommand{\fnum@table}{\sf\textbf{\tablename~\textbf{S}\textbf{\thetable}}}
\makeatother

\setcounter{figure}{0}
\setcounter{table}{0}
\setcounter{equation}{0}

\section*{Appendix}

\subsection*{Photograph classification process}

In order to ensure that only photographs representing an interaction between an individual and an ecosystem are considered, each photo has been manually validated and classified conforming to the Common International Classification of Ecosystem Services \cite{Haines2013}. We relied on this typology to identify and classify socio-ecological interactions according to different types of cultural services that people benefits from ecosystems. For each case study, the interpretation of the photographs was performed by between 1 and 6 local experts. Photos that were not relevant included the following categories: a) wrong geographic location; b) people or pets as main subject in the foreground, not representing an outdoor activity; c) indoor, parking, private gardens; d) vehicles in the foreground; e) objects, signs and logos not related to the landscape; f) photo duplicate; g) bad photo where the subject cannot be identified. The photos were classified according to different categories: aesthetic enjoyment of landscapes (wide views of natural or different kind of environments), recreational activities (e.g. photographs of sport activities, such as skiing, hiking, climbing, camping), aesthetic enjoyment or existence of species (photographs of animals or plants), or intellectual experiences such as education, artistic inspiration or cultural heritage (e.g. photographs of scientific field work, traditional livestock feeding practices, lifestyle related to agricultural heritage).

More specifically, the subject of each photo was manually validated and classified according to the landscape identified in the picture. We used six landscape categories: agricultural and open landscape, sparse forest landscape, forested landscape, mountain landscape, anthropic infrastructures, water landscapes and wetlands. At the end of the process, $16,716$ photos taken by $2,967$ users between January 2000 and 2017 were classified. Note that 98\% of the photos were taken after 2007. 

\subsection*{Identification of the user's place of residence}

To identify the place of residence of the $2,967$ Flickr users, we retrieved through the Flickr API information related to all the geo-located photos taken by these users worldwide. Then, we divided the world using a grid composed of $100 \times 100$ square kilometers cells in a cylindrical equal-area projection. We only considered photos with the most precise spatio-temporal Flickr accuracy level intersecting the grid and (Figure S\ref{FigS1}). For each cell visited by a user we count the number of distinct months during which at least one photo was taken from this cell. The place of residence of a user is given by the cell in which the user was present the higher number of months. The identification of the user's place of residence process allows us to discard non reliable Flickr users (collective account or not regular Flickr user). A first coarse filter was applied to exclude collective accounts from the data by filtering out users traveling faster than a plane ($750$ km/h). Then, to ensure that a user shows enough regularity and that the assigned place of residence is the region of the world where he/she is really living, we applied two filters. We considered only users having more than $N=6$ distinct months with at least one photo taken and a rate of presence at the place of residence higher than $\delta=1/3$. Where $\delta$ is the ratio between the number of distinct months with at least one photo taken in the cell of residence and $N$. These values represents a good trade-off between being relatively sure about the users' residence area and keeping enough number of users to have proper statistics (Figure S\ref{FigS2}). The algorithm used to extract most visited locations from individual spatio-temporal trajectories is detailed in \cite{Lenormand2016} and the source code is available online\footnote[1]{\url{https://www.maximelenormand.com/Codes}}. The final number of users per site is displayed in Figure S\ref{FigS3}.

\subsection*{Accessibility and attractiveness}

For each user $u$, we compute the distance $d_{us}$ between their place of residence, represented by the average position of all the photos taken from his/her cell of residence, and the centroid of the study site $s$ he or she has visited represented by the average position of all photos taken in the site by all users. However, this distance between user's origin and visited site can be biased by the geography. Indeed, some case study sites are more isolated than others, implying differences in terms of accessibility among sites. To take this heterogeneity into account, we define a measure of accessibility $\lambda_s$ as the average distance between the place of residence of every inhabitants on earth (estimated with the Global Human Settlement Population grid \cite{GHSL}) to the study sites. Hence, for every users $u$ that have visited at least once the site $s$ we can compute a normalized distance $\hat{d}_{us}$ taking into account the origin of $u$ and the accessibility of $s$ (Equation \ref{dus}). All the distances have been computed with the Haversine formula based on longitude and latitude coordinates. 
\begin{equation}
\hat{d}_{us}=\frac{d_{us}}{\lambda_s} \label{dus}
\end{equation}
The normalized distance is comprised between 0 and 3, $\hat{d}_{us}=0.015$ corresponds roughly to a distance of $100$ km (Figure S\ref{FigS4}). To ease interpretation the results are expressed in kilometers multiplying $\hat{d}_{us}$ by a factor $100/0.015$. Some statistics about the study sites and their accessibility is presented Table S\ref{Tab1}.

\begin{table}[!h]
	\caption{\textbf{Summary statistics of the case study sites.}}
	\label{Tab1}
	\begin{center}
		\begin{tabular}{lcc}
			\hline
			\centering Site & Surface (km$^2$)  &  Accessibility (km) \\
			\hline
			Barcelona  & 7,822 & 7,092 \\
			Cairngorms & 3,253 & 7,176 \\
			Carpathians & 326 & 6,187 \\
			Costa Vicentina & 895 & 7,701\\
			Danube & 5,782 & 6,083 \\
			De Cirkel & 181 & 6,866 \\
			Dovre & 2,271 & 6,830 \\
			French Alps & 255 & 6,896 \\
			Kainuu & 24,438 & 6,467 \\
			Kiskunsag & 1,720 & 6,379 \\
			Loch Leven & 97 & 7,169 \\
			Sierra Nevada & 3,657& 7,429 \\
			Stubai valley & 265 & 6,662 \\
			Trnava & 270 & 6,443 \\
			Vinschgau & 491 & 6,688 \\
			Warwickshire & 2,256 & 7,120 \\
			
			\hline
		\end{tabular}
	\end{center}
\end{table}

\subsection*{Interactive web application}

\begin{figure}[!ht]
	\centering 
	\includegraphics[width=\linewidth]{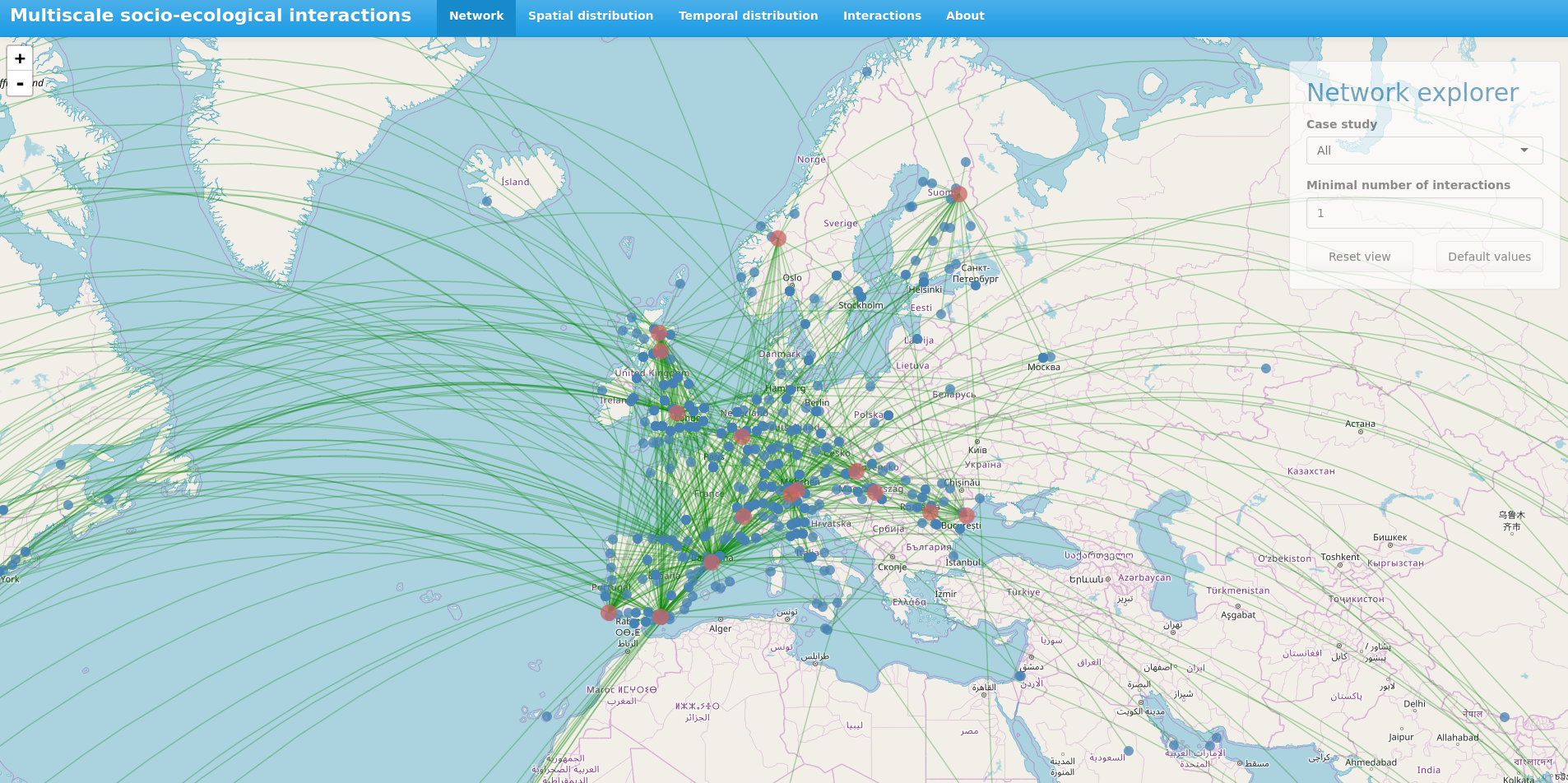}
	\caption{\textbf{Screenshot of the interactive web application.} \label{FigS0}}
\end{figure}

\begin{figure}[!ht]
	\centering 
	\includegraphics[width=\linewidth]{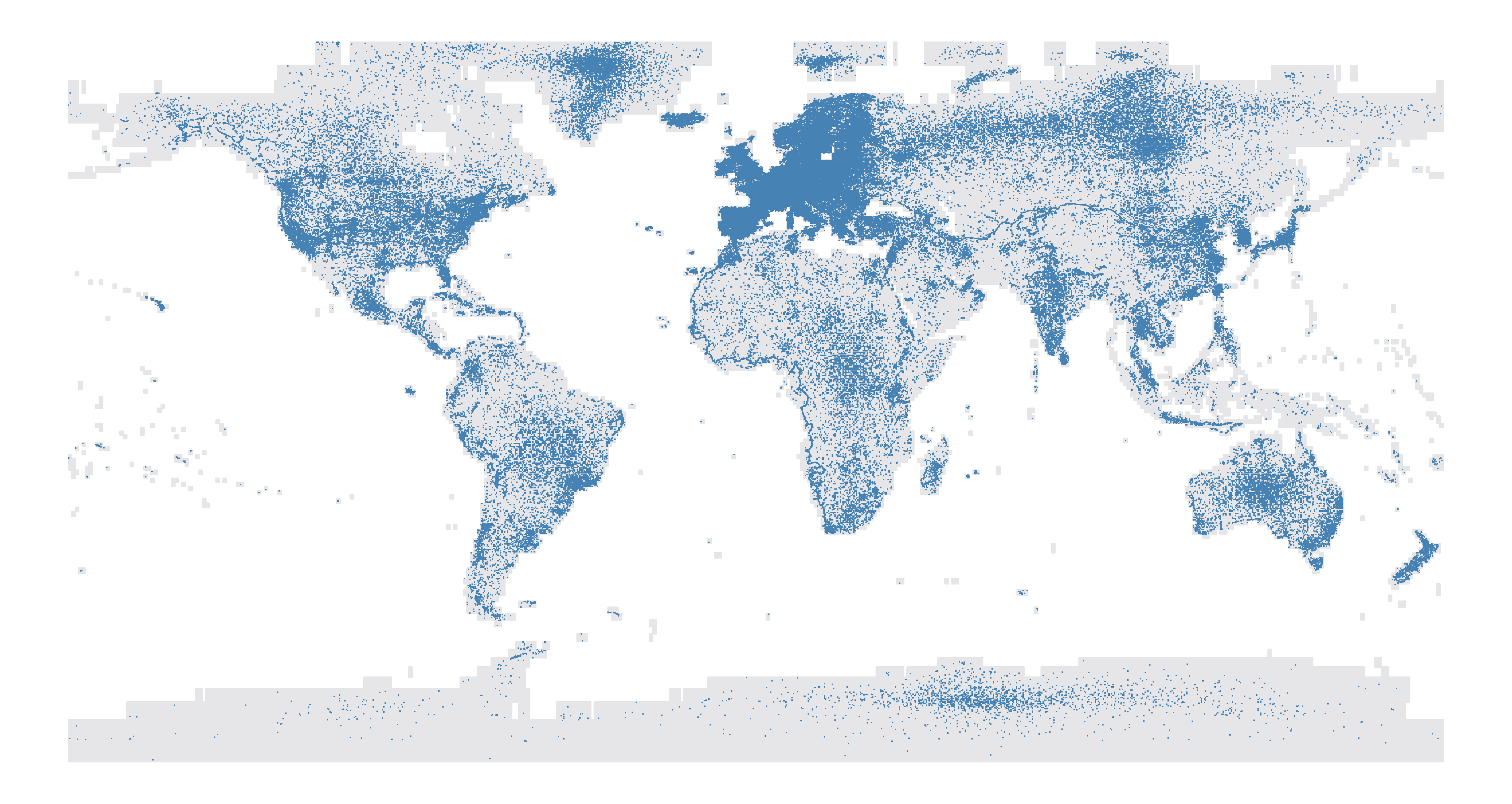}
	\caption{\textbf{Positions of the geolocated Flickr photographs.} Each photo is represented as a point on the map location from which it was taken. Then, we divided the world using a grid composed of $100 \times 100$ square kilometers cells in a cylindrical equal-area projection. We only considered the $5,353,356$ photos intersecting the world grid composed of $100 \times 100$ square kilometers cells in a cylindrical equal-area projection (background). \label{FigS1}}
\end{figure}

\begin{figure}[!ht]
	\centering 
	\includegraphics[width=15cm]{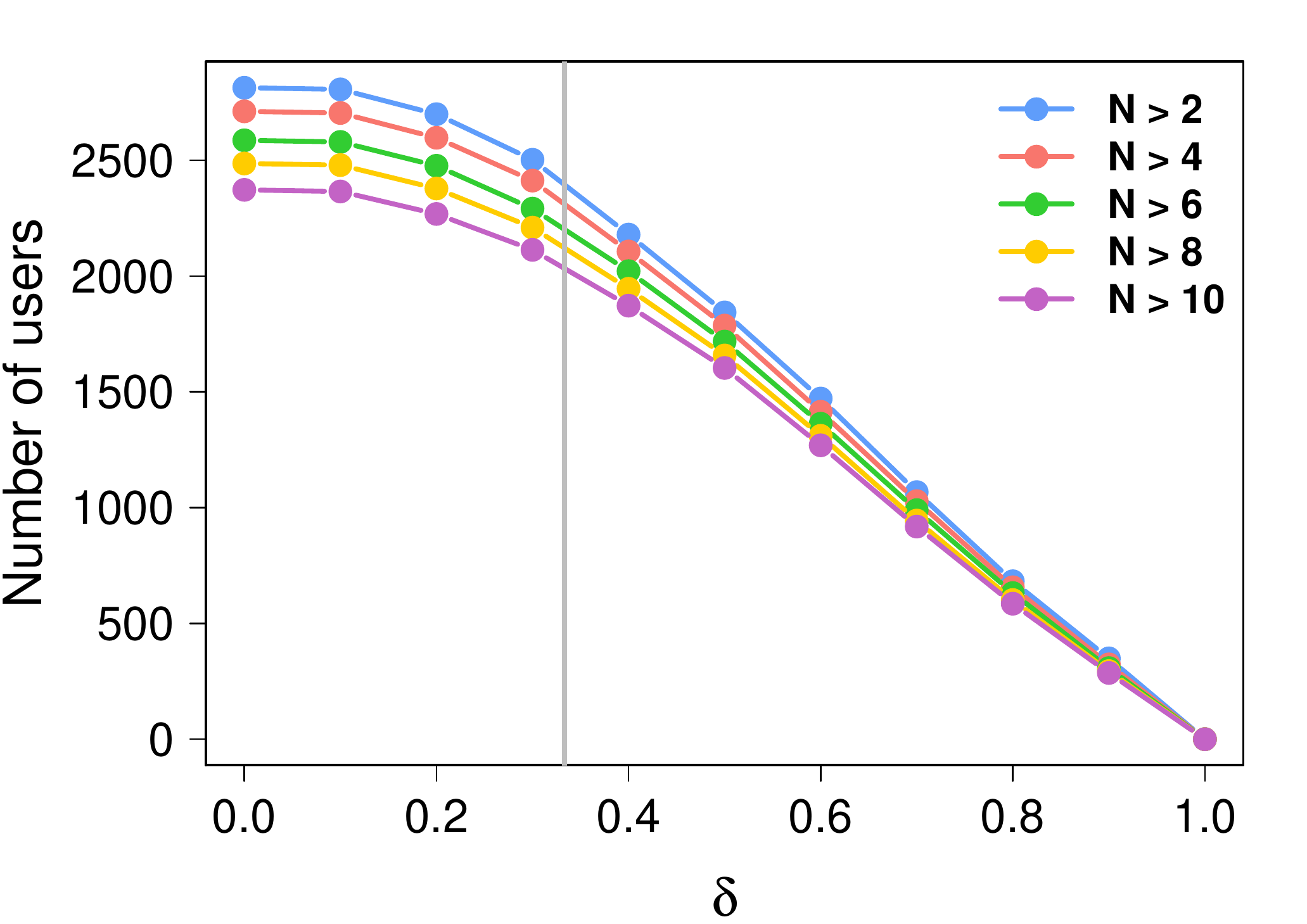}
	\caption{\textbf{Influence of the parameters in the identification of the Flickr user's place of residence.} Number of reliable users as a function of $\delta$ for different values of $N$. The vertical bars indicate the value $\delta=1/3$. \label{FigS2}}
\end{figure}

\begin{figure}[!ht]
	\centering 
	\includegraphics[width=15cm]{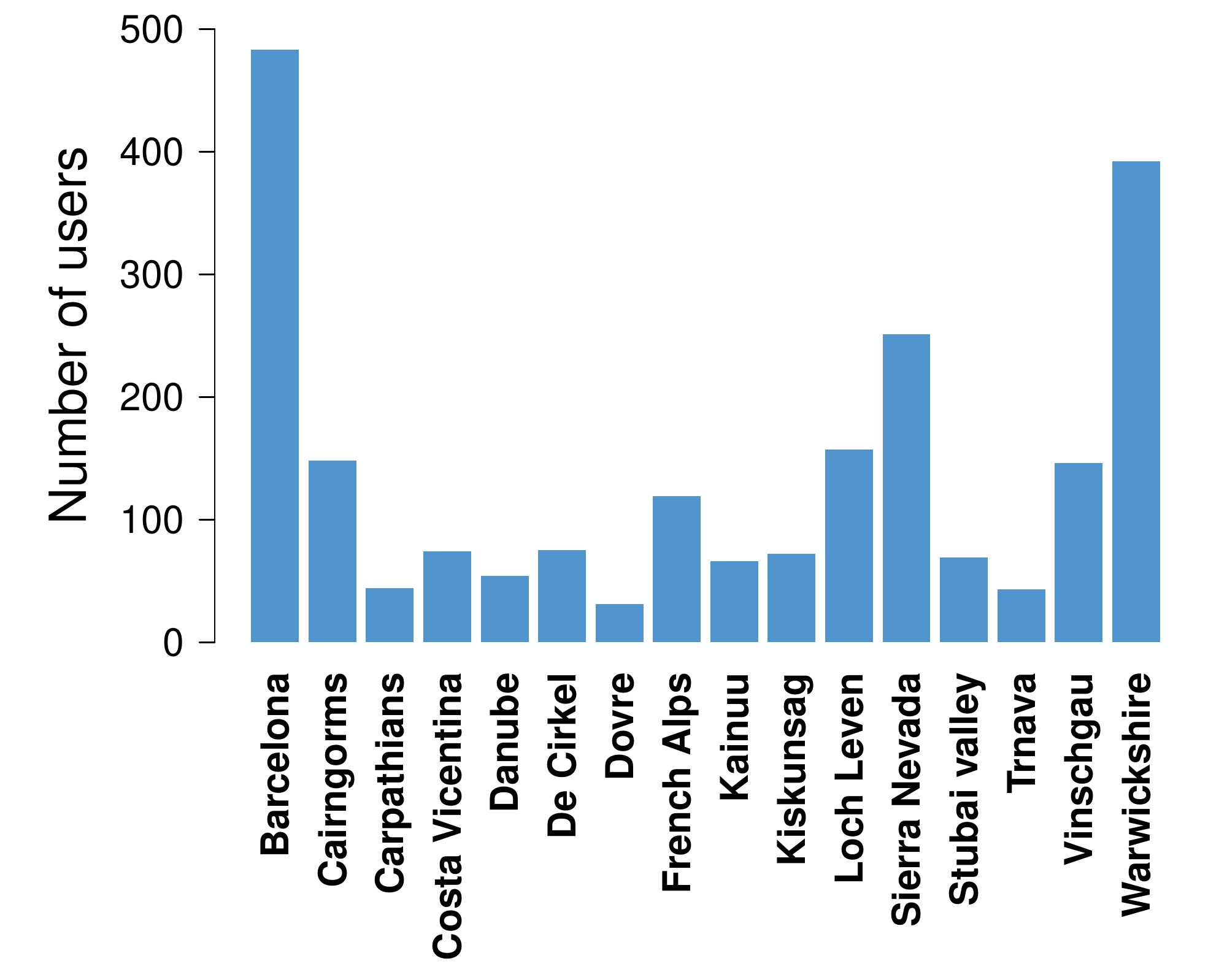}
	\caption{\textbf{Final number of users per site.} \label{FigS3}}
\end{figure}

\begin{figure}[!ht]
	\centering 
	\includegraphics[width=\linewidth]{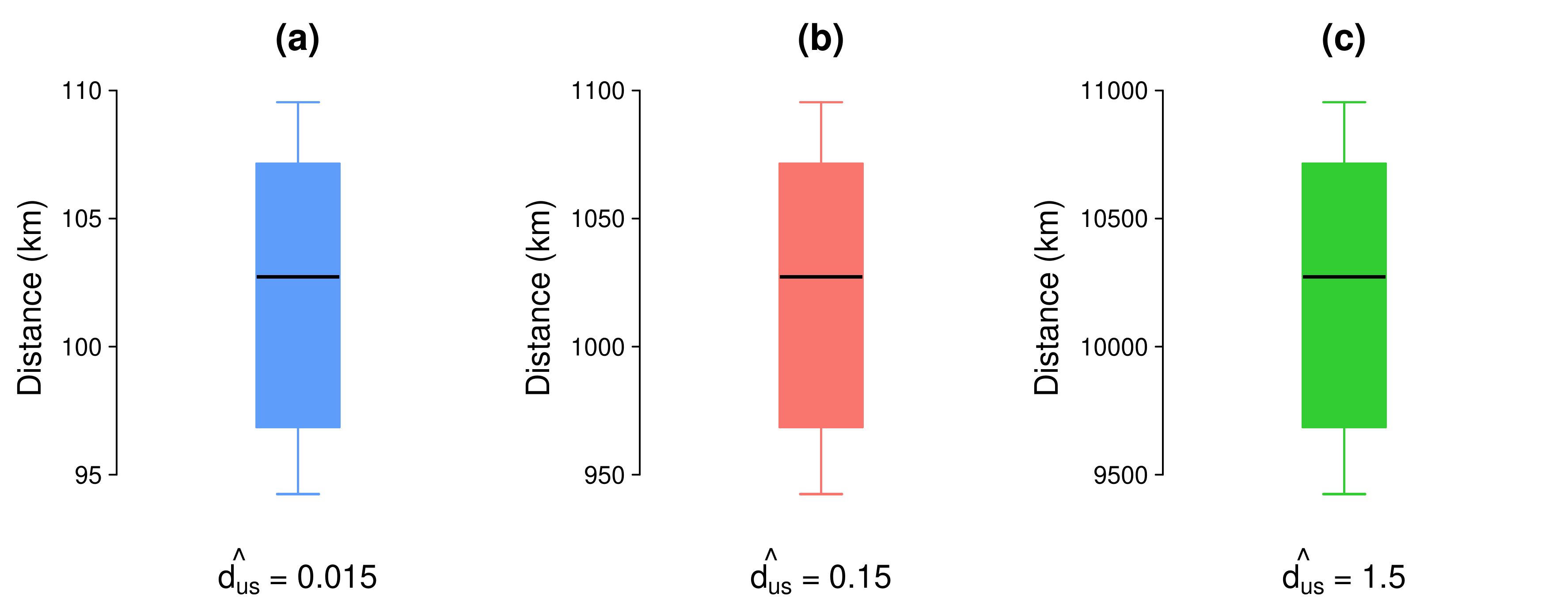}
	\caption{\textbf{Relationship  between the distance and the normalized distance across sites.} Boxplots of the distance from a site according to different normalized distance values $\hat{d}_{us}=0.015$ (a), $\hat{d}_{us}=0.15$ (b) and $\hat{d}_{us}=1.5$ (c).  The boxplot is composed of the first decile, the lower hinge, the median, the upper hinge and the 9$^{th}$ decile. \label{FigS4}}
\end{figure}

An interactive web application has been designed to provide an easy-to-use interface to visualize socio-ecological interactions at different scales in the 16 case studies across Europe (Figure S\ref{FigS0}). It was developed as part of a research project funded by the ALTER-Net network\footnote[2]{~\url{http://www.alter-net.info/}}. We focused on four aspects of the multiscale socio-ecological network: a representation of the spatial network at a world scale, a visualization of the spatial and temporal distribution of interactions per site, and, finally, a representation of the type of interactions (recreational activities and type of landscapes). The source code of the interactive web application can be downloaded from\footnote[3]{~\url{https://www.maximelenormand.com/Codes}}.

\onecolumngrid

\section*{Supplementary figures}

\begin{figure}[!ht]
	\centering 
	\includegraphics[width=\linewidth]{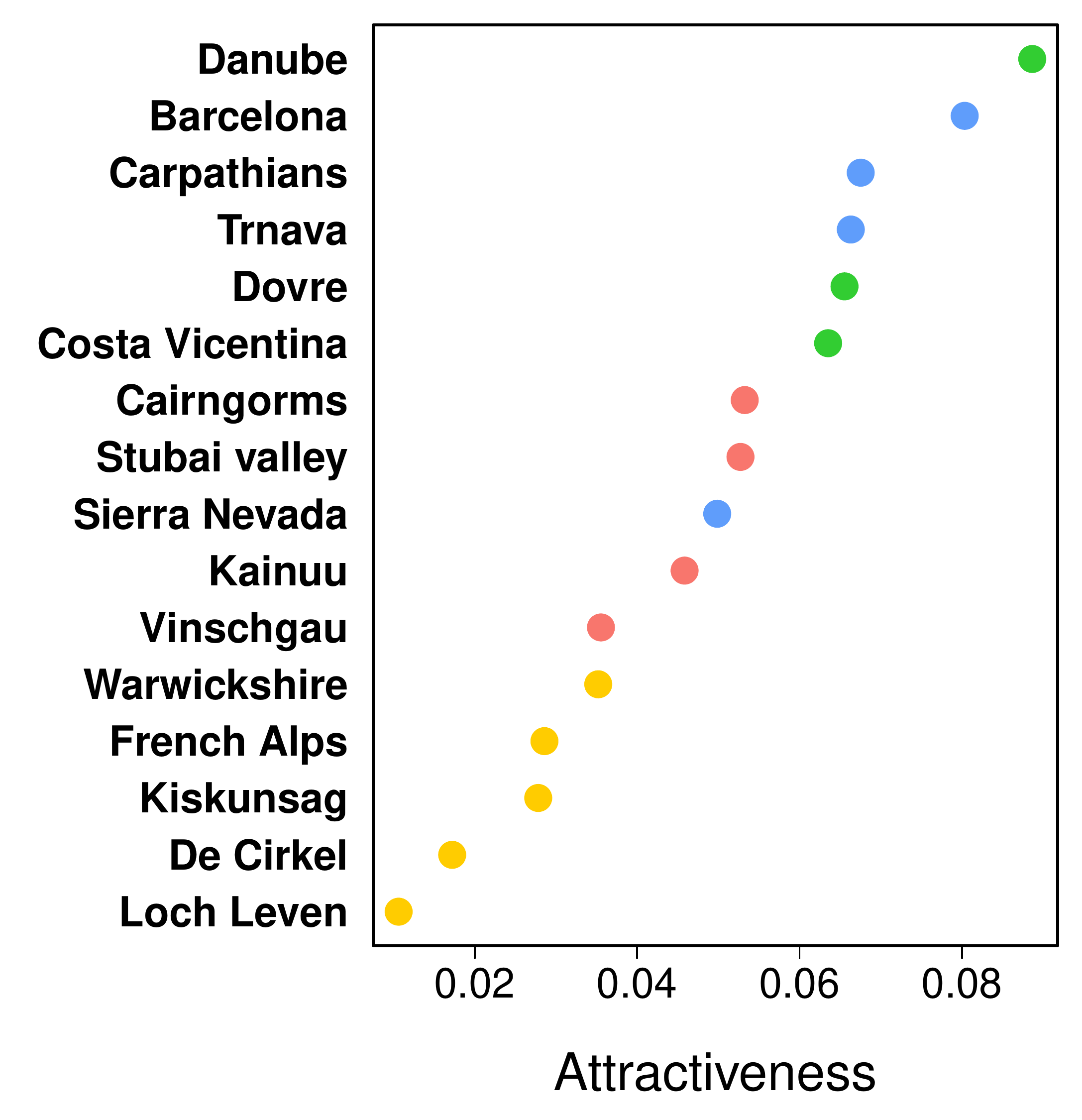}
	\caption{\textbf{Rankings of the case study sites according to their level of attractiveness.} The attractiveness of a site is equal to the area above the CDFs and the colors corresponds to the cluster analysis presented in Figure 5. \label{FigS5}}
\end{figure}

\begin{figure}[!ht]
	\centering 
	\includegraphics[width=\linewidth]{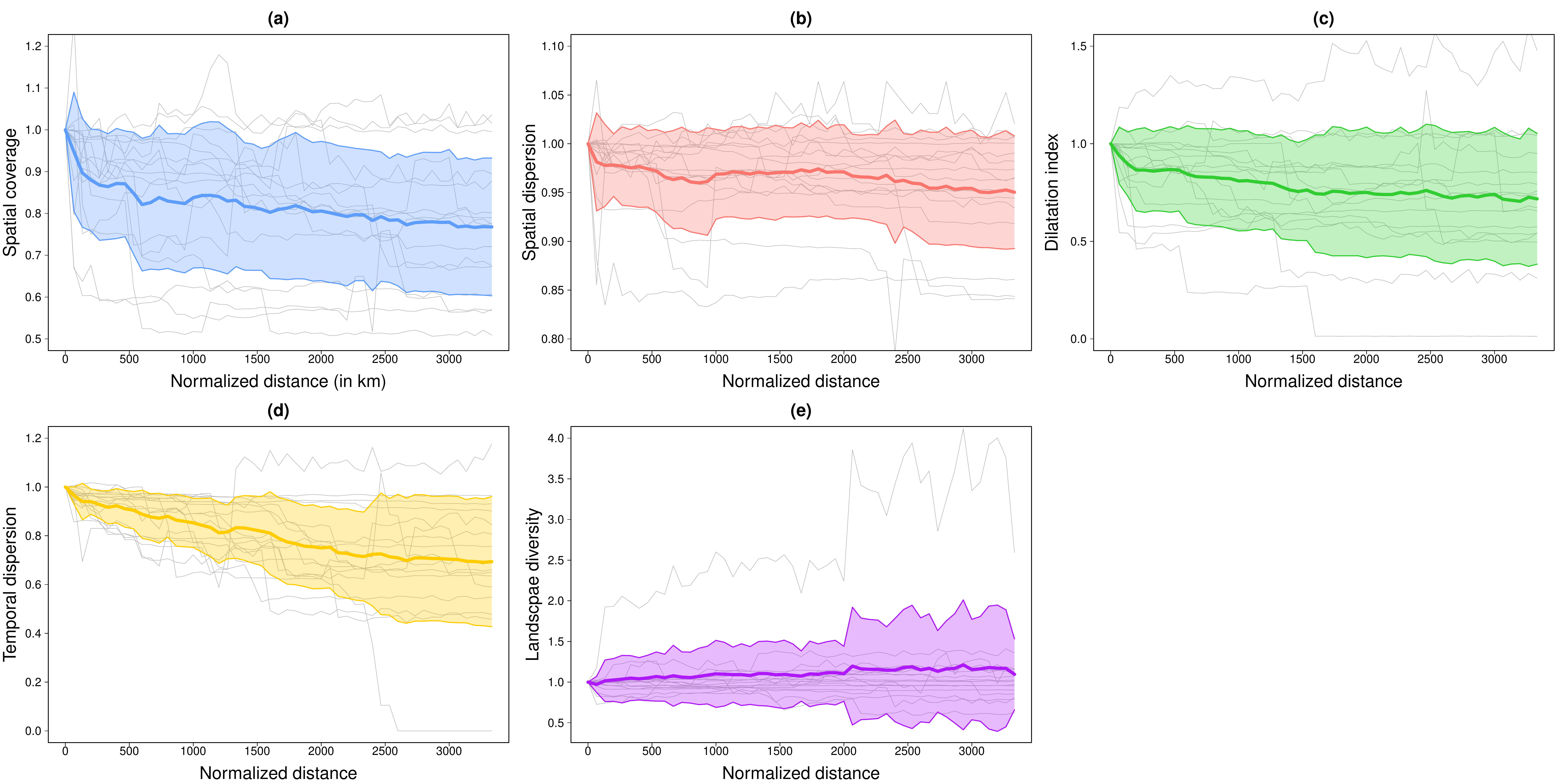}
	\caption{\textbf{Effect of the distance traveled on the socio-ecological interactions.} Evolution of the spatial coverage (a), the spatial dispersion (b), the spatial dilatation index (c), the temporal dispersion (d) and the landscape diversity (e) as a function of the normalized distance. Each grey curve represents a case study. For each metric, the mean and standard deviation over the 16 case studies are displayed. All metrics are normalized by the value obtained with the null model. \label{FigS6}}
\end{figure}

\begin{figure}[!ht]
	\centering 
	\includegraphics[width=\linewidth]{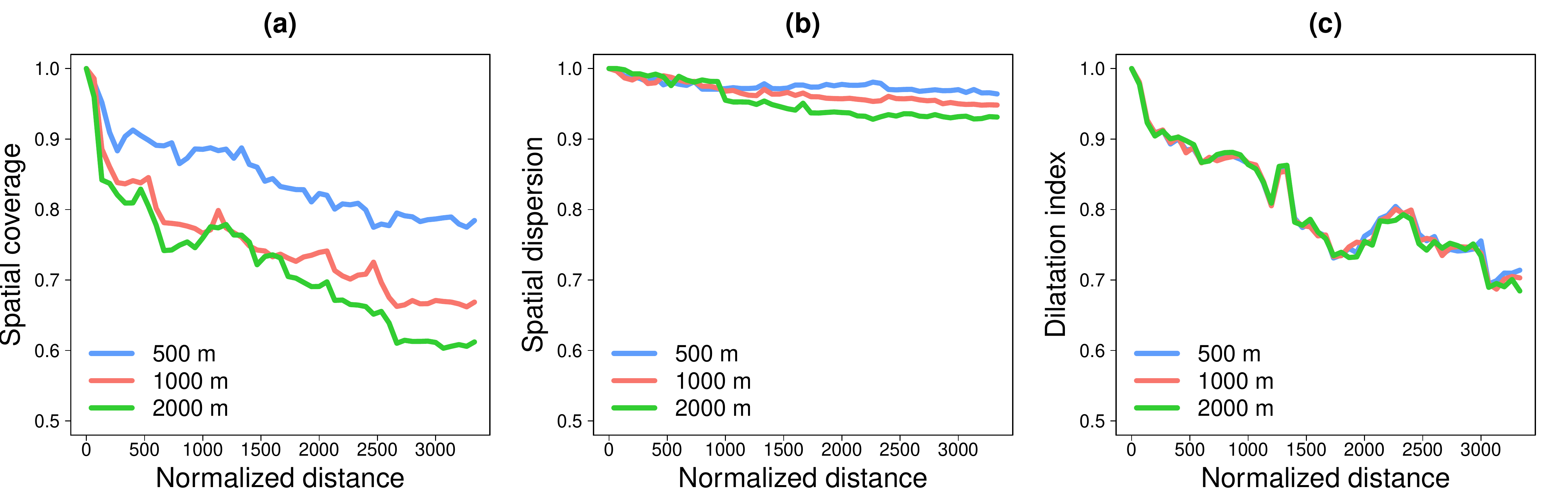}
	\caption{\textbf{Effect of the spatial granularity on the metrics.} Evolution of the spatial coverage (a), the spatial dispersion (b), the spatial dilatation index (c) as a function of the normalized distance according to the cell size (500, 1000 and 2000 meters). For each metric, the median over the 16 case studies is displayed. All metrics are normalized by the value obtained with a random null model. \label{FigS7}}
\end{figure}

\begin{figure}[!ht]
	\centering 
	\includegraphics[width=\linewidth]{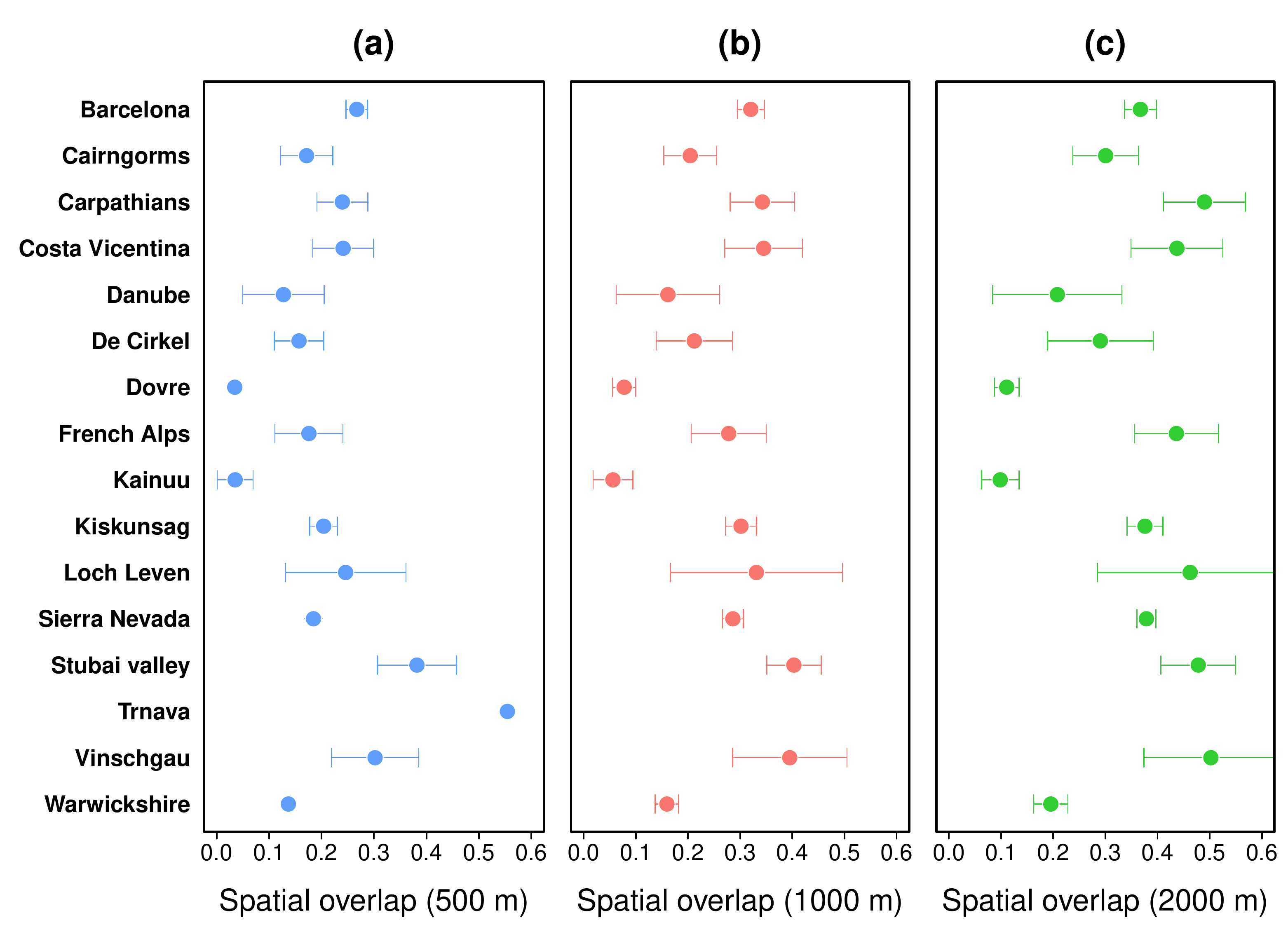}
	\caption{\textbf{Effect of the spatial granularity on the spatial overlap  between locals and visitors' interactions.} Different cell sizes are considered: (a) 500 meters; (b) 1000 meters; (c) 2000 meters. Locals and visitors are identified according to the normalized distance. In order to assess the impact of the threshold on the results we averaged the metrics obtained with threshold values ranging between 0.015 and 0.15. The error bars represent one standard deviation. \label{FigS8}}
\end{figure}

\end{document}